\def \AAP #1 #2 {{\em Astron. Astrophys.\/} {\bf #1}, #2}
\def \AAL #1 #2 {{\em Astron. Astrophys. Lett.\/} {\bf #1}, L#2}
\def \AAR #1 #2 {{\em Astron. Astrophys. Rev.\/} {\bf #1}, #2}
\def \AAS #1 #2 {{\em Astron. Astrophys. Suppl. Ser.\/} {\bf #1}, #2}
\def \AJ #1 #2 {{\em Astron. J.\/} {\bf #1}, #2}
\def \ANNREV #1 #2 {{\em Ann. Rev. Astron. Astrophys.\/} {\bf #1}, #2}
\def \APJ #1 #2 {{\em Astrophys. J.\/} {\bf #1}, #2}
\def \APJL #1 #2 {{\em Astrophys. J. Lett.\/} {\bf #1}, L#2}
\def \APJS #1 #2 {{\em Astrophys. J. Suppl.\/} {\bf #1}, #2}
\def \APSS #1 #2 {{\em Astrophys. Space Sci.\/} {\bf #1}, #2}
\def \ASR #1 #2 {{\em Adv. Space Res.\/} {\bf #1}, #2}
\def \BAIC #1 #2 {{\em Bull. Astron. Inst. Czechosl.\/} {\bf #1}, #2}
\def \JSQRT #1 #2 {{\em J. Quant. Spectrosc. Radiat. Transfer\/} {\bf #1}, #2}
\def \MN #1 #2 {{\em Mon. Not. R. Astr. Soc.\/} {\bf #1}, #2}
\def \MEM #1 #2 {{\em Mem. R. Astr. Soc.\/} {\bf #1}, #2}
\def \PLR #1 #2 {{\em Phys. Lett. Rev.\/} {\bf #1}, #2}
\def \PASJ #1 #2 {{\em Publ. Astron. Soc. Japan\/} {\bf #1}, #2}
\def \PASP #1 #2 {{\em Publ. Astr. Soc. Pacific\/} {\bf #1}, #2}
\def \NAT #1 #2 {{\em Nature\/} {\bf #1}, #2}
\def \SAIT #1 #2 {{\em Mem.\ Soc.\ Astron.\ It.\/} {\bf #1}, #2}
\def \MESS #1 #2 {{\em The Messenger\/} {\bf #1}, #2}
\def \ASTRNACH #1 #2 {{\em Astron. Nach.\/} {\bf #1}, #2}

\documentstyle{BSAXPwork}
\input epsf.sty
\begin{opening}
\title{Recent Discoveries of Energetic Young Radio Pulsars}
\author{D.R. Lorimer}
\institute{University of Manchester, Jodrell Bank Observatory, Macclesfield,
Cheshire, SK11 9DL, UK}
\date{} 
\end{opening}

\begin{document}

\oddpagefooter{}{}{} 
\evenpagefooter{}{}{} 
\medskip  

\begin{abstract}
The observed population of young radio pulsars has grown substantially
in recent years due to a combination of large-scale surveys and
deep targeted searches. Many of the pulsars are associated with
supernova remnants and/or unidentified gamma-ray sources. This review
summarises the state of play and looks ahead to likely advances in
the near future.
\end{abstract}

\def\lapp{\ifmmode\stackrel{<}{_{\sim}}\else$\stackrel{<}{_{\sim}}$\fi} 
\def\gapp{\ifmmode\stackrel{>}{_{\sim}}\else$\stackrel{>}{_{\sim}}$\fi} 

\section{Introduction}

\begin{figure}[hbt]
\epsfysize=5cm 
\epsfbox{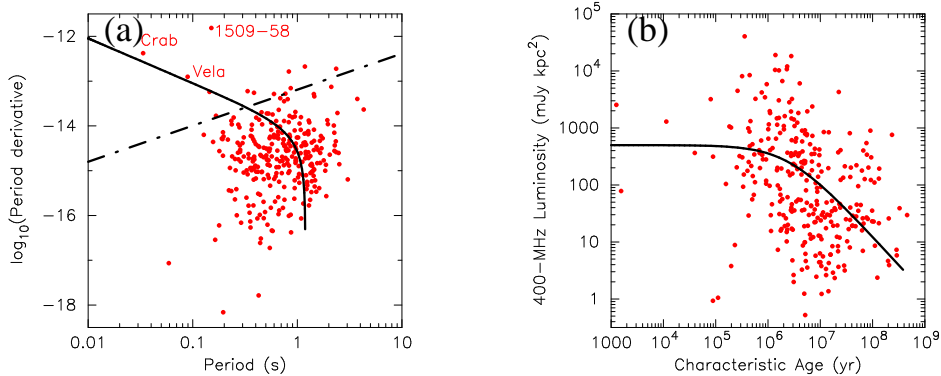}
\caption{\sl (a): The $P-\dot{P}$ diagram for the sample of $\sim 300$ 
pulsars known as of 1983. The solid line shows model spindown
evolution for 15 Myr assuming magnetic dipole braking with exponential
decay of the magnetic field with a 5-Myr time constant. The dashed
line represents a characteristic age of $10^5$ yr.
(b): 400-MHz luminosity versus characteristic age for the same
sample. The solid line shows luminosity evolution predicted by the
Lyne, Manchester \& Taylor (1985) model, again assuming exponential
magnetic field decay with a 5-Myr time constant.}
\label{fig:1983}
\end{figure}

The insightful prediction of Baade \& Zwicky (1934) that neutron stars
are produced during core-collapse supernovae gained significant
observational credence in 1968 with the discovery of young,
rapidly-rotating radio pulsars in the Crab and Vela supernova remnants
(Staelin \& Reifenstein 1968; Large, Vaughan \& Mills 1968). The
characteristic ages\footnote{The characteristic age
$\tau=P/(2\dot{P})$, where $P$ is the pulse period and $\dot{P}$ the
rate of slowdown.}  of these pulsars are respectively $\sim 10^3$ and
$10^4$ yr.  Fifteen years later, in 1983, despite the sample rising to
over 300 objects, only a further 9 young pulsars\footnote{Throughout
this review, ``young pulsars'' are those with characteristic ages less
than $10^5$ yr. This cutoff is roughly consistent with the maximum
observable age of a supernova remnant.} were known.
Fig.~\ref{fig:1983}a shows the 1983 sample in the period--period
derivative ($P-\dot{P}$) plane. Since we see only a snap-shot of the
population, most observed pulsars tend to be middle-aged ($10^{6-7}$
yr) objects with $\sim 0.5$-s spin periods (the ``island'' feature in
Fig.~\ref{fig:1983}a).

The rarity of young pulsars means that the first ones to be discovered
will likely be luminous. This is demonstrated in Fig.~\ref{fig:1983}b,
a plot of radio luminosity against characteristic age for the 1983
sample which shows an apparent dearth of young faint objects. Whether
this plot shows evidence for luminosity decay with age is 
debatable.  Although some luminosity decay might be expected a-priori
(for example if the magnetic field decays; Fig.~\ref{fig:1983}b), a
dangerous conclusion to draw from this plot is that young pulsars are
necessarily bright objects. Recent discoveries show that this
conclusion is incorrect.  Luminous young pulsars are the exception
rather than the rule.

The sample of associations between young neutron stars and supernova
remnants has grown substantially in recent years as a result of
observational progress in many areas.  In the latest census, Kaspi \&
Helfand (2002) tabulate 52 examples of associations\footnote{Many 
are candidate associations, requiring further observational 
evidence to confirm/refute them.} between young
neutron stars and supernova remnants.  The fact that radio pulsations
have been detected in only 18 (35\%) of the neutron stars so far 
highlights the importance of observations throughout the
electromagnetic spectrum to obtain a true picture of the population.
Issues such as the pulsar birth rate, the supernova rate, the fraction
of supernovae that produce a neutron star, the radio pulsar beaming
fraction, the birth properties of pulsars (luminosities, periods,
velocities etc.) and the subsequent spindown evolution (magnetic field
decay, braking index evolution etc.) can all be constrained by
studying a statistically significant sample of young objects. This
review looks at the astonishing progress in this field in the last few
years which has provided many insights into this traditionally elusive
part of the compact-object population.

\section{Finding young radio pulsars}

The two main ways to find young pulsars are blind surveys and targeted
searches.  In a blind survey, the aim is to cover a large area of sky
with good sensitivity with a reasonable investment of telescope time
and manpower.  The optimum region to find young pulsars is close to
the Galactic plane where their progenitors (the massive O and B stars)
are known to lie. In addition, supernova remnants and gamma-ray
sources likely to be associated with the pulsar population are also
found along the Galactic plane.

\begin{figure}[hbt]
\epsfysize=5cm 
\epsfbox{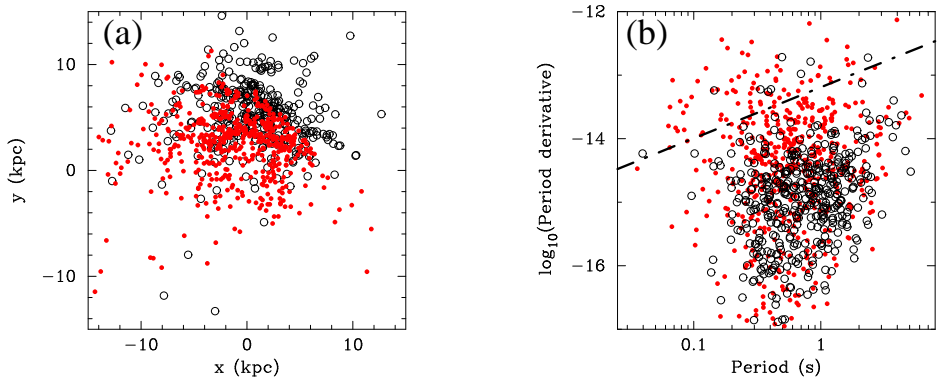}
\caption{\sl (a): The pulsar population projected onto the Galactic plane. 
In this projection, the Galactic centre lies at the origin and the Sun at 
(0.0,8.5). Open circles are pulsars previously known before the PM survey. 
Filled points represent new PM pulsar discoveries which lie predominantly 
in the inner Galaxy. (b): The same pulsar samples plotted in 
$P-\dot{P}$ space. As in Fig.~\ref{fig:1983}a, the dashed line is the
locus of points for which the characteristic age is $10^5$ yr.}
\label{fig:pksmb}
\end{figure}

By far the most successful project of this kind has been the Parkes
Multibeam (PM) survey of the Galactic plane which has so far
discovered over 600 pulsars (Manchester et al.~2001; Morris et
al.~2002; Kramer et al.~2003). The PM survey is about seven times more
sensitive than any previous Galactic plane search and, as shown in
Fig.~\ref{fig:pksmb}, has been particularly efficient at finding young
objects in the inner Galaxy.

In a targeted survey, the aim is to carry out the deepest searches
possible for radio pulsations by choosing well-motivated specific
directions in the Galaxy. For young pulsars, the obvious regions of
interest are supernova remnants and unidentified gamma-ray sources. In
the past, despite considerable effort, this approach was not very
fruitful (e.g.~Kaspi et al.~1996; Nice \& Sayer 1997; Lorimer et al.~1998). 
The main obstacle is the fact that most targets cannot be covered by a
single telescope beam (typically several arcminutes in extent). Since
there is often extreme competition for telescope time, searches
requiring a number of separate pointings are necessarily compromised
in sensitivity by the need to reduce the integration time to cover the
target area.

Targeted searches are now enjoying something of a renaissance with the
advent of new high-energy telescopes, such as {\it Chandra}. The
spatial resolution of the new instruments has resulted in
arcsecond localisation of neutron star candidates in a number of
supernova remnants. Armed with this information, and state-of-the art
receivers and data acquisition systems, observers at radio telescopes
have been able to carry out very deep single-pointing integrations
(often complete rise-to-set transits) of the high-energy point
sources, which has resulted in a dramatic leap in sensitivity.

\section{Pulsar properties inferred from their spindown}

Before moving onto the recent discoveries, let us summarise what can
be inferred about pulsar ages by studying their spindown.  Timing
measurements readily yield precise measurements of the pulsar rotation
frequency $\nu=1/P$ and its time derivative, $\dot{\nu}$. Both these
quantities provide insights into the age and spin history of the
neutron star. Expressing the spindown torque in general terms, we have
$\dot{\nu} \propto -\nu^n$, where $n$ is the so-called braking
index. Integrating this expression, for constant $n$, gives the pulsar age
\begin{equation}
\label{equ:age}
t = \frac{P}{(n-1)\dot{P}} \left[1 - \left(\frac{P_0}{P}\right)^{n-1}\right],
\end{equation}
where $P_0$ is the spin period at birth. Under the assumptions $n=3$
(spindown due to pure magnetic dipole radiation) and $P_0 \ll P$, this
expression reduces to $P/(2\dot{P})$, i.e.~the characteristic age,
$\tau$, defined earlier.

The young pulsar sample provides a means to investigate the validity
of the above assumptions a stage further.  To date, for five pulsars,
the high spindown rates allow measurement of $\ddot{\nu}$ which leads
to a direct determination of the braking index: $n=\nu
\dot{\nu}/\ddot{\nu}^2$. For these pulsars, braking indices
are in the range $1.4 < n < 2.9$, all below the expected value for
pure magnetic dipole radiation ($n=3$).  For pulsars where an
independent constraint on $t$ can be obtained (e.g.~from a convincing
association with a supernova remnant of known age), we may invert the
above expression to constrain the initial spin period.  For example,
for the Crab pulsar, where $n$ and $t$ are known, we find $P_0=19$ ms.

We shall discuss current constraints on $n$ and $P_0$ available from
the newly discovered pulsars in the following sections. Obtaining
reliable measurements of $n$ is non-trivial, since it requires a
careful treatment of the contaminating effects of so-called ``timing
noise'' prominent in young pulsars. Nevertheless, timing observations
currently underway have the potential to yield new measurements of
braking indices in the future.

\section{New pulsar/supernova remnant associations from the PM survey}

\begin{figure}[hbt]
\epsfysize=5cm 
\epsfbox{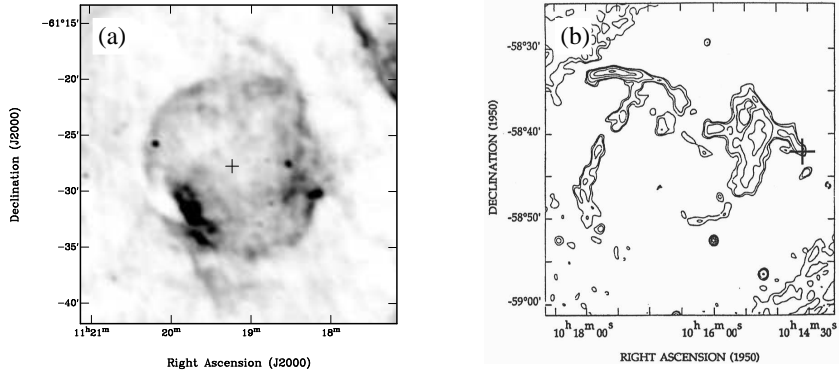}
\caption{\sl (a): G292.2--0.5 and PSR J1119--6127
(Crawford et al.~2001). (b): G284.3--1.8 and PSR 
J1016--5857 (Camilo et al.~2001). The pulsar positions
are marked by crosses.}
\label{fig:shells}
\end{figure}

PSR~J1119--6127 is a 407-ms pulsar discovered in the PM survey (Camilo
et al.~2000). With $\dot{P}=4\times10^{-12}$, currently the largest
known among radio pulsars, the characteristic age for this pulsar
$\tau = 1.6$ kyr. The extremely high rate of spindown permitted a
significant measurement of $\ddot{P}$ and hence a braking index of
$2.91\pm0.05$.  Inserting these parameters into equation (\ref{equ:age})
for any assumed value of $P_0$ implies an upper limit on the age of
this pulsar of only 1.7 kyr --- very similar to that of the Crab, B1509--58
and B0540--69, all of which are associated with supernova remnants.
Although no catalogued remnant was known at this position, a faint
shell was seen in the Molonglo 843-MHz survey (Green et
al.~1999). A dedicated Australia Telescope Compact
Array observation
revealed the non-thermal shell remnant G292.2--0.5 (Crawford et
al.~2001; Fig.~\ref{fig:shells}a).

The distance to the pulsar and remnant are currently not
well constrained (Camilo et al.~2000; Crawford et al.~2001). The line
of sight suggests that they may lie between two spiral arms in the
distance range 2.4--8 kpc. For a best-guess distance of 5 kpc,
Crawford et al.~argue the age of the supernova remnant to be less than
3 kyr. Another pointer to the youth and validity of the association is
the very small offset ($< 1'$) of the pulsar from the geometric centre
of the remnant. This is extremely unlikely to occur in a chance
alignment, especially with such a young pulsar, and implies that the
transverse velocity of J1119--6127 is less than 500 km s$^{-1}$. A
final constraint on the pulsar parameters comes from the lack of a
detection of a radio pulsar wind nebula, as is also the case in the
energetic pulsar B1509--58 (Bhattacharya 1990).  Extending
Bhattacharya's model to predict the brightness of a nebula around
J1119--6127, Crawford et al.~show that the lack of detection is
consistent with an initial spin period $<200$ ms. A distance estimate
to both the pulsar and remnant would further constrain the above arguments.

A second, albeit less certain, pulsar/supernova remnant association
from the PM survey, between PSR~J1016--5857 and G284.3--1.8 was
proposed recently by Camilo et al.~(2001).  As shown in
Fig.~\ref{fig:shells}b, the pulsar lies to the western edge of the
remnant, right at the tip of a bright ``finger'' of emission about $15'$
from the geometrical centre of the remnant. At the distance of
G284.3--1.8 (3 kpc; Ruiz \& May 1986) and assuming the 21 kyr
characteristic age of J1016--5857 to be correct, the implied pulsar
velocity, 500 km s$^{-1}$, is highly supersonic. Although no pulsar
wind nebula is obvious from the radio image, archival X-ray data from
the {\em Einstein Observatory} show a compact source in the 0.1--4.5
keV band at the pulsar position.  Camilo et al.~demonstrate that the
energetics of this source are consistent with a compact pulsar wind
nebula. Further multi-wavelength studies are required to investigate
the validity of the proposed association.

The above discoveries highlight the power of the blind pulsar survey
approach to find pulsar/supernova remnant associations.  Although
G292.2--0.5 is nominally above the detection threshold of supernova
remnant surveys, its previous non classification reflects the
difficulties in identifying remnants in crowded and complex regions of
the inner Galaxy. In addition to the above two cases, Manchester et
al.~(2002) discuss a further seven possible associations between
pulsars and supernova remnants in the PM survey. In two of these
cases, PSRs~J1726--3530 and J1632--4818, evidence for a supernova
remnant comes from a posteriori inspection of archival continuum
surveys. Further investigation of all these candidate associations is
now in progress.

\section{Recent progress in pulsar/EGRET source associations}

The nature of the $\sim 170$ unidentified sources from the EGRET
all-sky survey (see the most recent catalogue by Hartman et al.~1999)
has perplexed astronomers for some time. For the sources which have
been convincingly identified so far, the only counterparts are neutron
stars (for example the Crab, Vela and Geminga pulsars), the large
Magellanic cloud, a solar flare and AGN.  These
identifications have motivated many efforts to search the unidentified
sources for radio pulsar counterparts.  In the past, despite intensive
radio searches (see e.g.~Nice \& Sayer 1997), only a few convincing
associations between young energetic radio pulsars and the
$\gamma$-ray emission were found. Part of the problem was that the large
error boxes of the EGRET sources demanded a number of separate
radio-telescope pointings to fully cover the area. More recently,
however, progress has been made in finding new candidate associations
with radio pulsars through the numerous discoveries in the PM survey,
and also by careful targeted searches of several sources.

\subsection{Pulsar/EGRET source pair statistics}

To date, the PM survey has found just under 30 pulsars that lie within the
boundaries of EGRET source error boxes (Camilo et al.~2001; D'Amico
et al.~2001; Kramer et al.~2003). Positional coincidence alone, of
course, is not sufficient to associate these pulsars since the density
of pulsars along the Galactic plane is large and the probability of
chance alignment is significant.  By cross-correlating all pulsars
currently known with the EGRET catalogue, Kramer et al.~(2003) have
been able to quantify the number of chance associations that might be
expected and conclude that $19\pm6$ associations are likely to be
real. In order to prove an association is real, a
detection of the pulsar period in the relevant EGRET photons is
required. Although this is possible for some of the pulsars with
well-known ephemerides (e.g.~the Crab), back extrapolation of
ephemerides of recently discovered young pulsars is more problematic due 
in part to the high level of timing activity. More importantly, 
many of the EGRET
sources are weak and there are not many photons available for folding!
Ultimately, $\gamma$-ray detections for many of the pulsar candidates
will have to await future more sensitive telescopes such as GLAST.

In the absence of a detection in the EGRET photons, it is possible to
evaluate the likelihood of a detection using other factors such as the
spindown energy loss rate $\dot{E} \propto \dot{P}/P^3$, EGRET source
variability, and gamma-ray production efficiency $\eta = 4 \pi f d^2
F_{\gamma}/\dot{E}$.  Here, $f$ is the gamma-ray pulsar beaming
fraction, $d$ is the distance to the pulsar and $F_{\gamma}$ is the
gamma-ray flux of the EGRET source. For an assumed beaming fraction of
1 sr, $\eta<20$\% for the 8 or so associations that can be regarded as
being secure. As a result, large $\eta$ values for proposed
associations may be used as evidence for them being merely chance
alignments. Kramer et al.~(2003) investigated the current list with
this in mind and conclude that there are now a further 8 proposed
associations which look likely. The current total of 16 either
definite or likely associations is in good agreement with expectation
of $19\pm6$ pairs based on statistical grounds presented above.

\subsection{Recent discoveries from targeted searches}

The young pulsar~J2021+3651 was discovered by Roberts et al.~(2002)
during a campaign to find young pulsars in EGRET error boxes. In order
to minimise the time spent searching the full extent of the error
boxes, ASCA point sources with significant flux above 1 GeV (Roberts,
Romani \& Kawai 2001) were selected as targets for deep
observations. From a total of 5 sources selected and observed from
Parkes and Arecibo, radio pulsations were detected from just one
source, AX~J2021.1+3651, which lies within the EGRET source
3EG~J2021+3716. PSR~J2021+3651, with a period of 103 ms and spindown
rate of $9.56 \times 10^{-14}$, has very similar spin parameters to
the Vela pulsar, and is therefore most likely to be a young object
which will exhibit period glitches in the future. That the pulsar is
associated with 3EG~J2021+3716 seems likely, since the only other
X-ray source in the field is a Wolf Rayet star. Currently, the
distance to this pulsar is not very well constrained ($\gapp 10$ kpc)
from the pulsar dispersion measure (371 cm$^{-3}$ pc; rather large for
this line of sight), so that $\eta \gapp 15$\%. Currently, the
proposed association is very intriguing.  Further observations are
required to constrain the distance to the pulsar, and establish
whether an associated supernova remnant exists.

Another EGRET error box in the northern sky, 3EG~J2227+6122, has been
the subject of a multi-wavelength study by Halpern et al.~(2001a).
After ruling out many possible sources, Halpern et al.~concluded that
the only possible counterpart to the EGRET source is the {\em Chandra}
point source embedded within a faint radio shell. The most likely
conclusion is that the EGRET/X-ray source is an energetic young
neutron star and the radio shell is the associated pulsar wind nebula.
Follow-up observations of the X-ray source using the 76-m Lovell
telescope revealed PSR~J2229+6114, a faint 51.6-ms pulsar with a
characteristic age of only 10 kyr (Halpern et al.~2001b). Kothes,
Uyaniker \& Pineault (2002) have pointed out that the pulsar lies
at the edge of the supernova remnant G106.3+2.7, a larger feature
which contains the nebula found by Halpern et al.
The energetics of this nebula (which Kothes et al.~call the ``Boomerang'' 
nebula) are consistent with the energy output of J2229+6114.

Based on the initial radio ephemeris for J2229+6114, Halpern et
al.~(2001b) were able to detect X-ray pulsations in archival ASCA data
from 1999.  Although there is still considerable uncertainty, X-ray
data suggest the distance to the pulsar is $\sim 3$ kpc. Assuming 1 sr
beaming, the required efficiency of $\gamma$-ray production, $\eta
\sim 0.2$\%.  While the evidence for the association is compelling,
confirmation of the identification of PSR~J2229+6114 with
3EG~J2227+6122 would be a detection as a $\gamma$-ray pulsar. Although
a possible EGRET detection of PSR~J2229+6114 has recently been
reported by Thompson et al.~(2002), the significance level is low
($<3\sigma$).

\section{New pulsar/supernova remnant associations from {\em Chandra}}

The discovery of PSR~J2229+6114 was one of the first real indications
that young energetic pulsars may not be particularly radio-bright. The
1400-MHz flux density of this pulsar ($\sim 0.25$ mJy) is well below
the thresholds of most large-scale surveys. As a result, many faint
young pulsars remain to be discovered. Motivated by this, and the
discovery of {\em Chandra} point sources in several supernova
remnants, a number of deep radio searches for pulsations have been
carried out since mid 2001.

\subsection{PSR~J1124--5916 and G292.0+1.8}

The first new pulsar followed the discovery by Hughes et al.~(2001) of
a {\em Chandra} point source in the supernova remnant
G292.0+1.8. Targeting this position for 9.3 hr with the Parkes
telescope, Camilo et al.~(2002a) discovered the 135-ms pulsar
J1124--5916. Like PSR~J2229+6114, J1124--5916 is a weak radio source
with 1400-MHz flux density of $\sim 80 \mu$Jy. At the distance of
G292.0+1.8 (5 kpc; see discussion in Camilo et al.~2002a for details),
the luminosity in this band is only $\sim$ 2 mJy kpc$^2$.  The
association of J1124--5916 with the X-ray source has been confirmed
with the detection of 135-ms X-ray pulsations in a recent {\em
Chandra} observation (Hughes, private communication).

The characteristic age of J1124--5916, 2900 yr, is in good agreement
with a recent analysis by Gonzalez \& Safi-Harb (2003) of the original
Hughes et al.~(2001) {\em Chandra} observation of G292.0+1.8. Gonzalez
and Safi-Harb find an age of 2400--2900 yr under the assumption of that
the remnant is in a Sedov expansion phase. If this is correct, then it
implies that J1124--5916 was born spinning rapidly with a braking
index close to three. This latest age constraint is somewhat at
variance with an earlier esimate of 1700 yr from optical observations
(Murdin \& Clark 1979).  It is not clear to me (a humble radio
observer) which of these to believe.  Reconciling the pulsar age with
the Murdin and Clark estimate requires the initial pulsar spin period
to be $>90$ ms, or the braking index $n<3$. A future measurement of a
braking index for J1119--6127 would clarify this situation.  In
practice this may be difficult due to the large amount of timing noise
observed in this pulsar (Camilo, private communication).

\subsection{PSR~J0205+6449 and 3C58}

In the case of the Crab-like supernova remnant 3C58, {\em Chandra}
observations by Murray et al.~(2002b) identified a point X-ray source
and found 65-ms pulsations from it. A reanalysis of archival RXTE data
by Murray et al.~confirmed the existence of the new X-ray pulsar
J0205+6449 in archival RXTE data and showed that the spindown rate
$\dot{P} = 1.9 \times 10^{-13}$ so that the characteristic age $\tau
\sim 5400$ yr. This spindown rate was confirmed with the recent
discovery of extremely faint (0.5 mJy kpc$^2$) radio pulsations using
the Green Bank Telescope (Camilo et al.~2002b).

As historical evidence suggests that 3C58 is the remnant of the supernova
seen in the year 1181 (see e.g.~Stephenson \& Green 2002), the
age of 3C58 is only 822 yr. Murray et al.~reconcile the discrepancy
between this and the pulsar characteristic age by appealing to an
initial spin period of 60 ms for PSR~J0205+6449.  Other authors (see
for example Bietenholz, Kassim \& Weiler 2001) have argued, on the
basis of expansion measurements, that the age of 3C58 is in fact
closer to 5000 yr and that this is {\it not} the remnant of the SN
1181. As Murray et al.~comment, the larger age estimate assumes 3C58
to be in free expansion, an assumption which which may not necessarily
be correct.  Further understanding of the expansion measurement is
required before we can say with certainty that 3C58 is the remnant of SN 1181.

\subsection{PSR~J1930+1852 and G54.1+0.3}

{\em Chandra} observations by Lu et al.~(2001) revealed an X-ray point
source in the Crab-like supernova remnant G54.1+0.3. Based on the
morphology and energetics of the system, Lu et al.~proposed that
G54.1+0.3 is powered by a young rapidly spinning pulsar, as is the
case for the Crab nebula. In an attempt to discover the putative
pulsar, Camilo et al.~(2002c) carried out a 3-hr observation with the
recently upgraded Arecibo telescope and found the 136-ms pulsar
J1930+1852.  Once again, this is a weak radio source with 1175-MHz
flux density of only 60 $\mu$Jy, faint enough to fall below the
detection threshold of a pre-upgrade Arecibo search of G54.1+0.3
(Gorham et al.~1996).

\begin{figure}[hbt]
\epsfysize=5cm 
\epsfbox{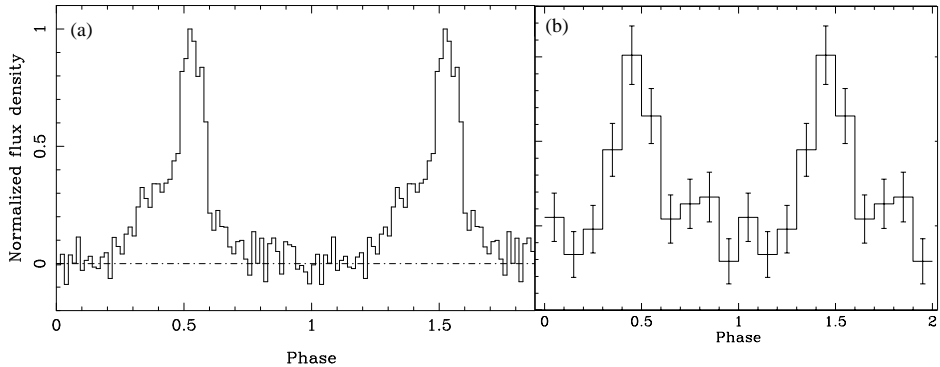}
\caption{\sl (a): Radio pulse profile of PSR~J1930+1852.
(b): ASCA X-ray profile. In both cases,
two pulse periods are shown and phase zero is arbitrary
(Camilo et al.~2002c).}
\label{fig:g54}
\end{figure}

Unfortunately, the time resolution of the original {\em Chandra}
observation was only 3 s (Lu et al.~2001). However, using the measured
spindown rate from the Arecibo observations, Camilo et al.~were able
to detect the 136-ms pulsations from archival ASCA data taken in
1999. This detection puts the association of the pulsations with the
X-ray source beyond doubt and cements PSR~J1930+1852 as the pulsar
powering G54.1+0.3.  Coincidentally, the spin parameters for
J1930+1852, and resulting characteristic age (2900 yr), are almost
identical to those for J1124--5916 discussed earlier. Unlike
G292.0+1.8, there is currently no independent age estimate for
G54.1+0.3. Arecibo timing observations of J1930+1852 currently
underway are aimed at absolute alignment of radio and X-ray profiles
and, timing noise permitting, a measurement of the braking index.

\section{PSR~J1747--2958: a young pulsar powering the ``Mouse'' nebula}

\begin{figure}[hbt]
\epsfysize=5cm 
\epsfbox{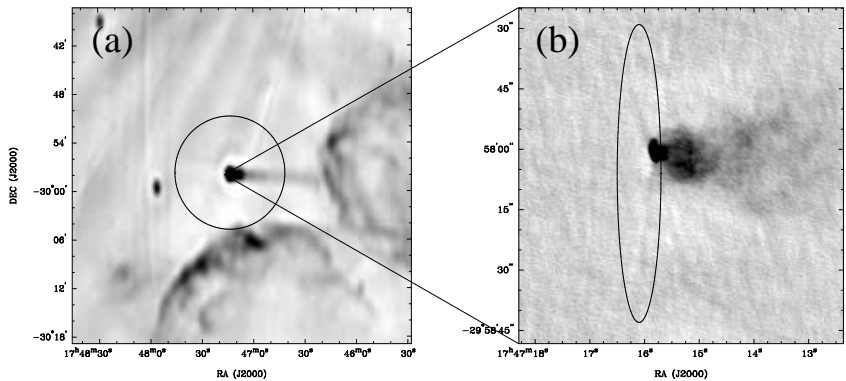}
\caption{\sl (a): MOST image of the Mouse Nebula at 0.8 GHz.
(b): 8.4-GHz VLA image.
The error ellipse shows the pulsar timing position uncertainty 
(Camilo et al.~2002d).}
\label{fig:mouse}
\end{figure}

The Mouse\footnote{Just where the term ``mouse'' comes from is a mystery;
based on the observed morphology of G359.23--0.82, the ``Tadpole
nebula'' seems a far more appropriate name!} nebula, G359.23--0.82, is an
extended non-thermal radio source discovered by Yusaf-Zadeh \& Bally
(1988) during a radio continuum survey with the VLA. As seen in the
radio images in Fig.~\ref{fig:mouse}, the Mouse is a point-like
feature with a wake of emission, highly suggestive of a compact source
moving supersonically through the local interstellar medium. ROSAT
observations by Predehl \& Kulkarni (1995) detected a point source
coincident with the head of the radio source and suggested that the
X-ray source is a young high-velocity pulsar powering the surrounding
radio nebula through its relativistic wind emission.

The putative pulsar, J1747--2958, was recently found in a 9.3-hr
observation of the Mouse using the Parkes telescope (Camilo et
al.~2002d). With $P=98$ ms and $\dot{P}=6 \times 10^{-14}$, the
characteristic age of 25 kyr strongly suggests that this is a young
pulsar. Although the pulsar position is not yet pinned down in
declination (the error ellipse shown in Fig.~\ref{fig:mouse}b has been
reduced as a result of further timing observations since this figure
was produced; Camilo, private communication), the probability of
finding a young pulsar this close to the tip of the Mouse by chance is
less than $5 \times 10^{-5}$.  It therefore seems likely that
J1747--2958 is the associated radio pulsar. From the pulsar dispersion
measure, and an independent distance constraint to the Mouse ($< 5$
kpc; Uchida et al.~1992), we follow Camilo et al.~(2002d) and take as
a best estimate of $\sim 2$ kpc for the distance to the Mouse.

Proceeding under the assumption that J1747--2958 is causing a
bow-shock nebula as it ploughs through the local interstellar medium,
we can make an estimate of the pulsar velocity, $V$, required to produce
the observed nebula. As discussed by a number of authors (see
e.g.~Chatterjee \& Cordes 2002), the stand-off radius $R_0$ between
the pulsar and the head of the bow shock scales with velocity as $R_0
\propto (\dot{E}/V^2)^{1/2}$, where the constant of proportionality
depends on the density of the local medium. Camilo et al.~make a quick
estimate of $R_0$ based on the angular separation from the high
resolution image shown in Fig.~\ref{fig:mouse}, for an assumed
distance of 2 kpc (see above). The pulsar velocity can then be
parameterised as $V \sim 570$ km s$^{-1}/\sqrt{n}$, where $n$ is the
(unknown) local medium number density in units of cm$^{-3}$.

The key measurement in this system in the near future will be the
pulsar proper motion, currently predicted at 60 mas yr$^{-1}$ for 570
km s$^{-1}$ at a distance of 2 kpc. A measurement of this quantity
(e.g.~using {\em Chandra} to perform the astrometry), will not only
confirm the association of the pulsar and nebula (from the direction
of proper motion), but will also constrain the age of the
system. Currently, a crude estimate of the age comes from the length
$L$ of the tail of emission in Fig.~\ref{fig:mouse}. Given the pulsar
velocity $V$, the age can be written as simply $t = L/V \sim 12
d_2/V_{570}$ kyr, where $d=2 \, d_2$ kpc and $V=570 \, V_{570}$ km
s$^{-1}$. A proper motion measurement would place a distance-independent
constraint on the age and birth spin period of PSR~J1747--2958.

\section{Future prospects}

Further discoveries of energetic young radio pulsars will undoubtably
be made in the near future as the PM survey continues, and high-energy
instruments continue to reveal neutron star candidates.  Although
these searches have been particularly successful, there are a number
of sources which have been searched for which there is, so far, no
known radio pulsar counterpart. One example is the famous Cas-A point
source discovered during a first-light {\em Chandra} observation
(Tannanbaum 1999).  To date, no statistically significant pulsations
have been found in X-ray searches (Murray et al.~2002a).  McLaughlin
et al.~(2001) placed a luminosity limit of 20 mJy kpc$^2$ on radio
pulsations this from a 2-hr observation with the VLA. Since this
observation does not preclude the existence of a faint pulsar akin to
those discovered in other {\em Chandra} sources, a group of us
(McLaughlin, Hankins, Kern and myself) have recently taken a much more
sensitive (18-hr) observation of the source with the VLA. Data
analysis is proving challenging due to the radio-frequency
interference environment at the VLA. At the very least, we are confident
of improving upon the upper limit once the analysis is complete.

Ultimately, a careful study of all the detections and upper limits
from the latest round of surveys should be able to place strong
constraints on the beaming fraction and luminosity distribution of
young pulsars, and test previous model predictions. For example, on
the basis of pulsar statistics about a decade ago, we postulated
(Lorimer et al.~1993) that few pulsars were born with 1400-MHz
luminosities\footnote{In quoting this number, I have scaled the
400-MHz luminosity quoted in the original paper (30 mJy kpc$^2$) to
1400 MHz assuming a radio spectral index of --1.6.} below 4 mJy
kpc$^2$.  The recent discoveries of young pulsars with luminosities
well below this limit show that this statement is probably incorrect,
and that the fraction of faint young pulsars is significant. It should
be possible to quantify this remark in the near future.

\acknowledgements
It is a pleasure to thank Maura McLaughlin, Michael Kramer and
Fernando Camilo for their comments on the manuscript, as well as Froney
Crawford for providing Fig.~\ref{fig:shells}a. Many thanks also
to Fernando Camilo for numerous discussions about young pulsars, and recent
collaborative projects. My research is funded by the Royal Society.

\end{document}